\documentclass[acmsmall,authorversion]{acmart}

\usepackage{graphicx} 
\usepackage{subfig} 
\usepackage{array,varwidth,lipsum} 
\usepackage{multirow} 
\usepackage{pbox} 
\usepackage{tabularx}
\usepackage{xcolor,colortbl} 
\definecolor{Gray}{gray}{0.9}
 \newcolumntype{L}{>{\raggedright\arraybackslash}X}
\expandafter\def\expandafter\UrlBreaks\expandafter{\UrlBreaks\do\a%
\do\b\do\c\do\d\do\e\do\f\do\g\do\h\do\i\do\j\do\k\do\l\do\m\do\n%
\do\o\do\p\do\q\do\r\do\s\do\t\do\u\do\v\do\w\do\x\do\y\do\z\do\&}

\def\BibTeX{{\rm B\kern-.05em{\sc i\kern-.025em b}\kern-.08emT\kern-.1667em\lower.7ex\hbox{E}\kern-.125emX}}
    
\setcopyright{acmlicensed}
\acmJournal{PACMHCI}
\acmYear{2019} \acmVolume{3} \acmNumber{CSCW} \acmArticle{61} \acmMonth{11} \acmPrice{15.00}\acmDOI{10.1145/3359163}

\begin{document}

%

\title[Detection of Bots Spreading Religious Hatred in Arabic Social Media]{Hateful People or Hateful Bots? Detection and Characterization of Bots Spreading Religious Hatred in Arabic Social Media}

%

\author{Nuha Albadi}
\orcid{0000-0003-1273-8371}
 \affiliation{%
 \institution{Taibah University}
 \department{Department of Computer Science}
 \city{Medina}
 \country{Saudi Arabia}}
 \affiliation{%
 \institution{University of Colorado Boulder}
 \department{Department of Computer Science}
 \streetaddress{1111 Engineering Dr}
 \city{Boulder}
 \state{Colorado}
 \postcode{80309}
 \country{USA}}
 \email{nuha.albadi@colorado.edu}

\author{Maram Kurdi}
 \affiliation{%
 \institution{Taif University}
 \department{Department of Computer Science}
 \city{Taif}
 \country{Saudi Arabia}}
 \affiliation{%
 \institution{University of Colorado Boulder}
 \department{Department of Computer Science}
 \streetaddress{1111 Engineering Dr}
 \city{Boulder}
 \state{Colorado}
 \postcode{80309}
 \country{USA}}
 \email{maram.kurdi@colorado.edu}
 
\author{Shivakant Mishra}
\affiliation{%
 \institution{University of Colorado Boulder}
 \department{Department of Computer Science}
 \streetaddress{1111 Engineering Dr}
 \city{Boulder}
 \state{Colorado}
 \postcode{80309}
 \country{USA}}
\email{mishras@colorado.edu}

%
\renewcommand{\shortauthors}{N. Albadi et al.}

\begin{abstract}

Arabic Twitter space is crawling with bots that fuel political feuds, spread misinformation, and proliferate sectarian rhetoric. While efforts have long existed to analyze and detect English bots, Arabic bot detection and characterization remains largely understudied. In this work, we contribute new insights into the role of bots in spreading religious hatred on Arabic Twitter and introduce a novel regression model that can accurately identify Arabic language bots. Our assessment shows that existing tools that are highly accurate in detecting English bots don't perform as well on Arabic bots. We identify the possible reasons for this poor performance, perform a thorough analysis of linguistic, content, behavioral and network features, and report on the most informative features that distinguish Arabic bots from humans as well as the differences between Arabic and English bots. Our results mark an important step toward understanding the behavior of malicious bots on Arabic Twitter and pave the way for a more effective Arabic bot detection tools.

\end{abstract}


%

%
\maketitle
\section{Introduction}
The analysis of social media content to understand online human behavior has gained significant importance in recent years \cite{silva2016analyzing,magdy2016isisisnotislam,waseem2016hateful,mondal2017measurement}. However, a major limitation of the design of such analysis is that it often fails to account for content created by bots, which can significantly influence the messaging in social media. A social bot is an autonomous entity on social media that is typically engineered to pass as a human, often with the intent to manipulate online discourse~\cite{ferrara2016rise}. Recent studies have shown that a significant majority of the social media content is generated by bots. For example, a six-week study by the Pew Research Center found that around two-thirds of all tweets with URL links were posted by likely bots~\cite{Pewlinks}. As a result, the presence of bots can negatively impact the results of social media analysis and misinform our understanding of how humans interact within the online social space. In particular, any social media analysis that doesn't take into account the impact of bots is incomplete. While some bots can be beneficial (e.g., customer service chatbots), the focus in this work is on content-polluter bots that mimic human behavior online to spread falsified information~\cite{ratkiewicz2011truthy}, create a false sense of public support~\cite{bessi2016social}, and proliferate dangerous ideologies~\cite{benigni2017online,berger2015isis}. 

Bots have vigorously invaded online social communities. A recent study estimated that bots constitute about 15\% of all Twitter accounts ~\cite{varol2017online}. With 321 million Twitter accounts~\cite{Twitterusers}, the implication is that there are more than 48 million bot accounts on Twitter. Twitter reported that the number of bots and trolls suspended each week is on the rise reaching 9.9 million as of May 2018~\cite{Twitterbots}. While this number may seem promising, Twitter's fight against bots is far from over~\cite{Twitterfailing}. 

Detecting bots in social media is a first step to account for the impact of bots in social media analysis. Our interest is in analysis of abuse in Arabic Twitter space, specifically the spread of religious hate, and thus to account for the impact of bots in our research, this paper focuses on detecting Arabic Twitter bots that are active in spreading hateful messages against various religious groups. Detecting bots in social media is challenging as bot designers are using sophisticated techniques to make a social bot look and behave as close to a human as possible~\cite{ferrara2016rise}. Several researchers have looked at the problem of detecting bots in Twitter (See Section~\ref{BotDetection}), and several bot detection tools are freely available~\cite{davis2016botornot,varol2017online,Botcheck} providing fairly high detection accuracy. However, we show in this paper that these tools fail to perform as well on Arabic Twitter bots as they do on English Twitter bots. In fact, Arabic Twitter bot detection and analysis is a considerably under-researched area. A study by Abokhodair et al.~\cite{abokhodair2015dissecting} analyzed a Twitter botnet that was active during the Syrian civil war to understand how it might have influenced related discussions. El-Mawass et al.~\cite{el2016detecting} estimated that around 75\% of Saudi trending hashtags on Twitter contained spam content, some of which was created by automated spammers. 

In our recent work on hate speech in Arabic social media ~\cite{albadi2018they,Albadi2019}, we showed that Arabic Twitter is awash with religious hatred which we defined as ``a speech that is insulting, offensive, or hurtful and is intended to incite hate, discrimination, or violence against an individual or a group of people on the basis of religious beliefs or lack thereof". Having such a large volume of hate speech and knowing that ISIS and other radical organizations have been using bots to push their extreme ideologies~\cite{benigni2017online,berger2015isis}, we hypothesize that bots may be to blame for a significant amount of this widespread hatred. 


In this work, we build a novel regression model, based on linguistic, content, behavioral and topic features to detect Arabic Twitter bots to understand the impact of bots in spreading religious hatred in Arabic Twitter space. In particular, we quantitatively code and analyze a representative sample of 450 accounts disseminating hate speech from the dataset constructed in our previous work \cite{albadi2018they,Albadi2019} for bot-like behavior. We compare our assigned bot-likelihood scores to those of Botometer~\cite{davis2016botornot}, a well-known machine-learning-based bot detection tool, and we show that Botometer performs a little above average in detecting Arabic bots. Based on our analysis, we build a predictive regression model and train it on various sets of features and show that our regression model outperforms Botometer's by a significant margin (31 points in Spearman's rho). Finally, we provide a large-scale analysis of predictive features that distinguish bots from humans in terms of characteristics and behaviors within the context of social media. 

To facilitate Arabic bot detection research and Twitter automation policy enforcement, this paper provides the following findings and contributions. 
\begin{enumerate}
    \item We quantify the impact of bots in terms of percentages of hateful tweets and accounts in spreading religious hate on Arabic Twitter.
    
    \item We show that bot detection tools that are highly effective in detecting English Twitter bots are not as effective in detecting Arabic Twitter bots.
    
    \item We develop a highly accurate Arabic bot detection regression model, which takes into account linguistic, content, behavioral, and topic features.
    
    \item We identify a subset of features that are highly discriminatory in detecting Arabic bots and provide insights into what makes these features discriminatory.
    
    \item We make our dataset consisting of 450 manually-labelled accounts publicly available to researchers to facilitate future research in this and related areas\footnote{https://github.com/nuhaalbadi/ArabicBots}.
\end{enumerate} 

\section{Background and Related Work}
In this section, we first discuss the main challenges encountered in analyzing Arabic language and social media content in general. We then survey prior research on online hate speech and bot detection and analysis. 

\subsection{Challenges of Arabic Language and User-generated Content}
The Arabic language poses unique challenges to the process of analyzing and studying online content \cite{farghaly2009arabic,darwish2012language}. Arabic is a morphologically rich language with a substantial amount of syntactic and relation information encoded at the word level. Arabic is also a pluricentric language with varying dialects corresponding to various regions in the Arab world. Words can have entirely different meanings across dialects. Social media users tend to use Arabic dialects rather than Modern Standard Arabic (MSA). Unlike MSA, Arabic dialects are not standardized, and often fail to follow well-defined language rules and structures. Besides, Arabic is a greatly under-resourced language with few Natural Language Processing (NLP) tools supporting MSA, let alone Arabic dialects \cite{diab2004automatic}. 

Other challenges that are encountered while studying user-generated content include multilingual text, slangs, misspellings, abbreviations, and lengthening of words. Furthermore, microblogging platforms that impose a maximum length on posts such as Twitter can lead to text that lacks context, which in turn may lead to a flawed analysis. Moreover, some online users tend to mask abusive and hateful content by presenting it as a harmless joke or hiding it inside a comical image. Such behavior can lead to abusive and toxic content going undetected. We describe later in this paper how these aforementioned challenges have been addressed.  

\subsection{Online Hate Speech}
Our previous work ~\cite{albadi2018they,Albadi2019} appears to be the only one focusing on hate speech detection and analysis in Arabic social media. Our study revealed that religious hate speech is widespread on Arabic Twitter. We found that almost half of the tweets discussing religion preached hatred and violence against religious minorities, mostly targeting Jews, Atheists, and Shia (the second largest Islamic sect). In particular, we found that there was a 60\% chance that a tweet would be hateful if it contained the Arabic equivalent of the word Jews. 

 To provide a sense of comparison between the volume of hate speech on Arabic Twitter and English Twitter, we report the results of a study conducted by Magdy et al. ~\cite{magdy2016isisisnotislam}, in which they analyzed a large volume of English tweets mentioning Islam while reacting to the 2015 Paris attacks. Their analysis suggested that only 17\% of such tweets were directing hate toward Muslims, 61\% were spreading positive messages about Islam, while 22\% were natural. 

A growing body of hate speech research has been conducted on English social media content. Distinguishable among this work are studies related to the detection of online hateful content targeting race and gender using character $n$-grams ~\cite{waseem2016hateful}, word embeddings~\cite{badjatiya2017deep}, and document embeddings~\cite{djuric2015comment}. A measurement study conducted by Silva et al.~\cite{silva2016analyzing} exploring the main targets of hate speech on Twitter and Whisper, an anonymous social media platform, showed that black people were the most targeted group on both networks, followed by white people on Twitter and fake people on Whisper. While race was the main targeted category on Twitter, behavior (e.g., sensitive people) was the main targeted category on Whisper. 

\subsection{Malicious Use of Bots in Social Media}
While previous research has studied harmless bots on several collaborative and social platforms such as Wikidata \cite{hall2018bot}, Twitch \cite{seering2018social}, and Reddit \cite{kiel2017could}, our focus is on malicious bots. Previous studies have thoroughly investigated such nefarious roles that can be played by bots, particularly in English online social space. One of such roles is political {\it astroturfing} wherein a flood of bot accounts (usually created by a single entity) creates the illusion of public support for a particular political candidate for the purpose of influencing public opinion. Bessi and Ferrara~\cite{bessi2016social} suggested that social bots have generated about one-fifth of the 2016 U.S. Presidential election discourse on Twitter. Twitter confirmed this in an official blog post~\cite{TwitterRussian} reporting that approximately 1.4 million accounts were notified about having some form of interactions with suspicious Russian-linked accounts (trolls and bots) who were spreading misinformation during the 2016 U.S. election. This nefarious use of bots is not new to social media; Ratkiewicz et al.~\cite{ratkiewicz2011truthy} indicated that bots have been used to amplify fake news and misinformation during the 2010 U.S. midterm elections through a coordinated generation and liking of misguiding tweets. It has also been shown that bots are used by ISIS propagandists to inflate their influence on Twitter and popularize their extreme ideologies~\cite{benigni2017online,berger2015isis}.  

Limited research has been conducted to study bot behavior on Arabic social media. The only relevant research we are aware of is the work by Abokhodair et al.~\cite{abokhodair2015dissecting}, in which they analyzed a Syrian botnet consisting of 130 bots that were active for 35 weeks before being suspended by Twitter. Their analysis suggested that the main task of such bots was to report news from a highly biased news source. A different but related research problem is the detection of spam content which sometimes involves bots. In \cite{el2016detecting}, El-Mawass et al. reported that about 74\% of tweets in Saudi trending hashtags are spam. They suggested that bots are sometimes used to increase the reach of spam content by coordinated liking and retweeting of spam tweets.  

\subsection{Bot Detection}\label{BotDetection}
\begin{figure}[ht]
  \centering
  \includegraphics[width=\linewidth]{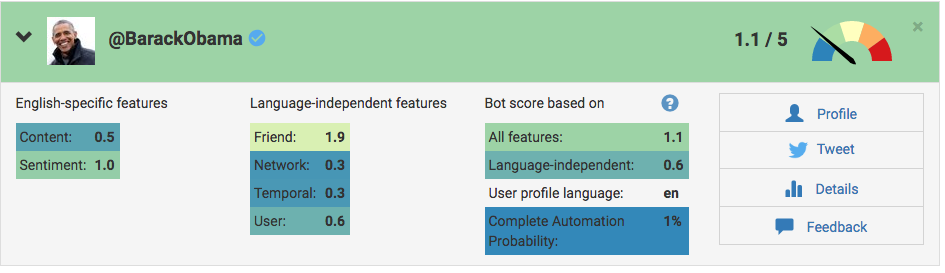}
  \caption{Example Botometer bot scores interface. }
  \label{Botometer}
\end{figure}

There are two main approaches to detecting social media bots in literature: supervised learning and unsupervised learning. An example of a supervised-based bot detection model is Botometer\footnote{\url{https://botometer.iuni.iu.edu}}~\cite{davis2016botornot,varol2017online}, which is a freely available tool that employs supervised machine learning algorithms to predict a bot score, from 0 to 5, for public Twitter accounts. This score indicates the likelihood of an account being a bot based on 1,150 features distributed across six feature categories. Botometer also computes an individual bot score for each of the six feature categories, comprised of friend features (e.g., local time and popularity of retweeting and retweeted accounts), network features (e.g., network metrics that describe distribution and density of retweet and mention networks), user features (e.g., number of followers, number of friends, profile language, account age), temporal features (e.g., average time between two consecutive tweets, tweeting rate), content features (e.g., length of tweet, frequency of part-of-speech tags), and sentiment features (e.g., arousal, valence, and dominance scores). Figure \ref{Botometer} provides an example of Botometer's bot score interface. It is worth noting that although content and sentiment features are computed for non-English tweeting bots, they are only meaningful for English tweeting bots. Botometer conveniently provides a language-independent bot score, which we considered in our study. 

DeBot~\cite{chavoshi2017demand}, on the other hand, utilizes unsupervised techniques to detect Twitter bots based on synchronicity and activity correlation between accounts. The system has several services that can answer the following questions. Is a given account a bot? How long has it been active? Which bots are currently tweeting about a given topic? Which bots are participating in a given hashtag? They compared their system to Botometer and found that 59\% of bots detected using their system had a Botometer bot score exceeding 50\% (Botometer's previous scoring scheme ranged from 0\% to 100\%). Their analysis suggested that bots in a given botnet share the same tweets 87\% of the time. 

To our knowledge, no existing work has attempted to specifically detect Arabic bots. In ~\cite{morstatter2016new}, Morstatter et al. created a dataset of 3,602 Arabic tweeting bots using a honeypot trap mechanism and a human dataset consisting of 3,107 users---a high bot ratio we argue that doesn't represent an actual bot percentage on Twitter, which is estimated to be between 9\% and 15\%~\cite{varol2017online}. Our work is different from ~\cite{morstatter2016new} in several important aspects. First, the main goal in ~\cite{morstatter2016new} is to improve recall in detecting bots while our goal is to detect Arabic bots with high precision in the context of religious hate. Second, Morstatter et al. created a binary classifier to classify whether an account is a bot or not; as bots nowadays are very sophisticated with many of them exhibiting both human and bot behaviors at the same time ~\cite{varol2017online}, we argue that the problem can't be simplified into a binary classification problem. To address this issue of mix behaviors, we adopt two techniques: instead of using any automated mechanism such as setting up a honeypot trap, we rely on manual labeling of accounts by assigning each a score ranging from 0 to 5 which indicates the degree of bot-like behavior an account is showing to get the ground truth; and we create a regression predictive model trained on our manually-labeled accounts to predict bot scores for new Twitter accounts. Finally, our work specifically focuses on the unique characteristics of Arabic bots, and thus provides deep insights into the predictive features that distinguish bots from humans and broadens the understanding of bots' behavior in the context of Arabic social space. 

\section{Data and Prior Analysis}
\subsection{Data Collection}
To identify accounts disseminating hate speech, we started working from the {\it hate speech} dataset constructed in our previous work ~\cite{albadi2018they,Albadi2019}, which consists of 6,000 Arabic tweets collected in November 2017 and annotated for religious hate speech by crowdsourced workers (see Table \ref{hatespeechDataset} for general statistics of the dataset). The tweets were collected by querying Twitter's Standard search\footnote{\url{https://developer.twitter.com/en/docs/tweets/search/api-reference/get-search-tweets}} API using impartial terms that refer to one of the six most common religious groups across the Middle East and North Africa. Although we didn't use offensive terms or religious slurs in our data collection process, the number of returned hateful tweets was surprisingly large. More details on the construction and analysis of this dataset can be found in ~\cite{albadi2018they,Albadi2019}. 


\begin{table}
  \caption{General statistics of the hate speech dataset \cite{albadi2018they,Albadi2019}.}
  \label{hatespeechDataset}
  \begin{tabular}{ll}
    \toprule
    Total tweets & 6,000\\
    Hateful tweets & 2,526 (42.1\%) \\
    Non-hateful tweets & 3,090 (51.5\%) \\
    Unrelated tweets & 384 (6.40\%) \\
    Hateful tweets referencing Jews & 598 (59.80\%)\\
    Hateful tweets referencing Atheists & 564 (56.40\%)\\
    Hateful tweets referencing Shia & 495 (49.5\%)\\
    Hateful tweets referencing Christians & 361 (36.10\%)\\
    Hateful tweets referencing Sunnis & 116 (11.60\%)\\
    Hateful tweets referencing Muslims & 24 (2.40\%)\\
  \bottomrule
\end{tabular}
\end{table}

In this dataset, we identified 4,410 unique Twitter accounts. Of these, 543 accounts were suspended, and thus we excluded them from our study. We then looked at the remaining 3,867 active accounts and classified them into accounts with hateful tweets or accounts with non-hateful tweets based on the number of hateful and non-hateful tweets they had authored. If they had authored more hateful tweets than non-hateful tweets, we classified them as accounts with hateful tweets. This resulted in having 1,750 accounts with hateful tweets and 2,117 accounts with non-hateful tweets. Since this study is focused on identifying the role of bots in spreading religious hatred, only accounts with hateful tweets were considered.  

For each account with hateful tweets, we collected up to 3,200 of their recent tweets using the GET statuses/user\_timeline\footnote{\url{https://developer.twitter.com/en/docs/tweets/timelines/api-reference/get-statuses-user\_timeline.html}} method from Twitter's API. The total number of collected tweets was more than 4.2 million tweets. We also collected each account profile information (e.g., location, time zone, language) using the GET users/show\footnote{\url{https://developer.twitter.com/en/docs/accounts-and-users/follow-search-get-users/api-reference/get-users-show.html}} API method. 

\subsection{Ground Truth}

To evaluate the accuracy of Botometer scores in Arabic Twitter, we need to get a ground truth of which accounts are bots. Getting the ground truth for such an inherently difficult task is not straightforward. Some of the approaches proposed in literature are fully automatic approaches without any manual inspection, e.g. setting a honeypot trap \cite{lee2011seven,morstatter2016new}, identifying synchronicity and correlation in accounts' behavior \cite{chavoshi2017demand}, and observing accounts getting suspended by Twitter \cite{morstatter2016new}. Others have relied on manual labeling of accounts into bot-like or human-like \cite{varol2017online}. The snowball mechanism has also been used in which researchers identify a seed list of highly suspicious accounts and then snowball their connections to collect a larger set of suspicious accounts \cite{wang2013social,abokhodair2015dissecting}.

The common aspect among all earlier efforts in labeling of bots is that they assign binary labels to accounts, bot or human. Given that there is no simple list of rules that can decisively identify bots, we argue that it is more effective to assign labels on a scale to reflect the inherent uncertainty in recognizing bots. Additionally, since modern bots that attempt to hide themselves are becoming more sophisticated, we argue that any fully automatic approach without any manual inspection to get ground truth about bots is bound to suffer from high inaccuracies.

Thus, we turn to manual labeling approaches to get the ground truth. Although crowdsourced workers can be helpful in many labeling and classification tasks, we argue that our task of fine-grained scoring of accounts on the level of bot-like behavior they are exhibiting requires a high-level of domain knowledge as well as extensive training that is hard to control in a crowdsource setting. We argue that in order get a reasonable set of ground truth data for identifying bots, manual labeling must be done by experts. Therefore, in order to insure high-quality labeling, the labeling of the accounts was done by two members of the research team who are native Arabic speakers and have gone through the following training steps to gain the required expertise to make a sound and informed judgment.

\begin{figure}[ht]
  \centering
  \includegraphics[width=10cm,height=10cm,keepaspectratio]{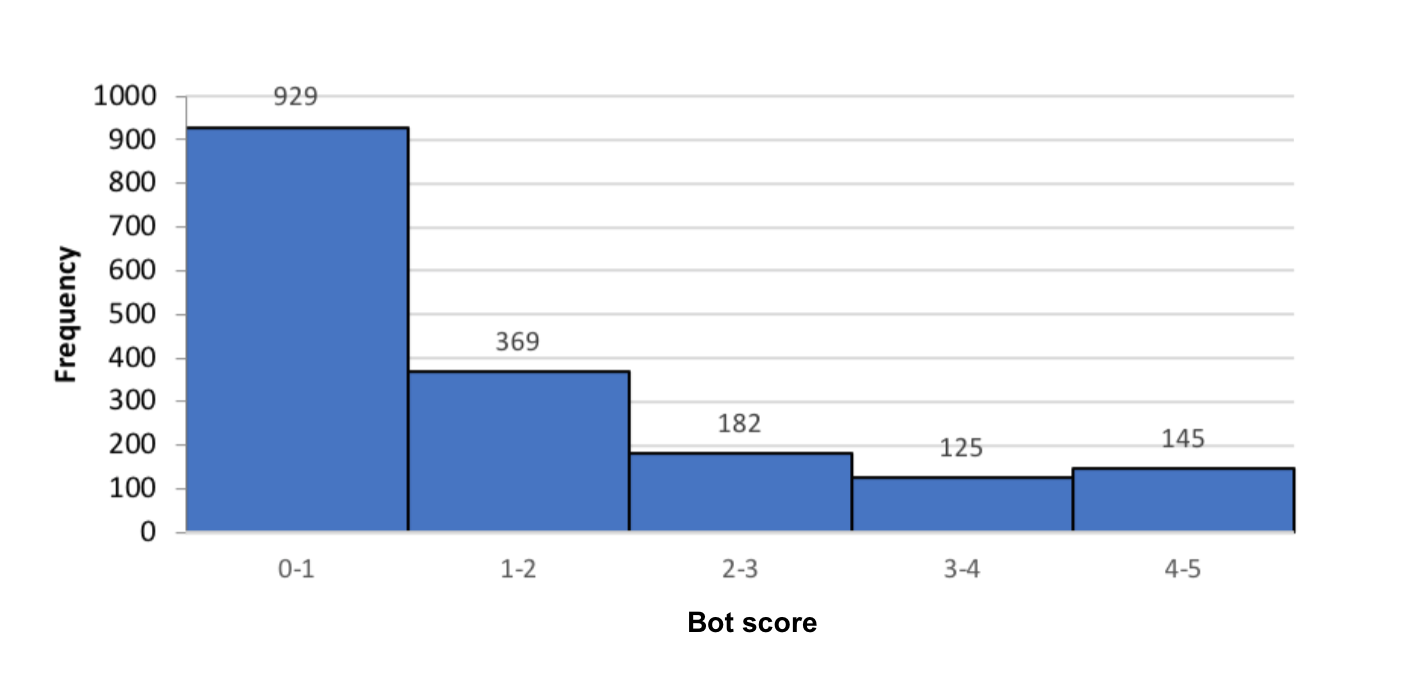}
  \caption{Botometer scores frequency distribution of accounts with hateful tweets.}
  \label{Botometer_scores}
\end{figure}

First, as a data exploration step, we applied Botometer on the 1,750 accounts with hateful tweets to discern the distribution of their bot scores (illustrated in Figure \ref{Botometer_scores}). Recall that bot scores from Botometer (we refer to this as Botometer scores) are on a scale from zero to five, with zero being ``most likely human'' and five being ``most likely bot''. A score in the middle of the scale is an indication that Botometer is unsure about the classification. As shown in this figure, the distribution is skewed to the right with the majority of accounts being assigned a Botometer score from 0 to 1.

Second, in order to gain the required domain knowledge with respect to bot behaviors and characteristics, we carefully examined the top 50 accounts receiving the highest Botometer scores as well as highly suspicious propaganda bots flagged by Botcheck.me~\cite{Botcheck}, a free online tool that is trained to identify English propaganda bots. We noted every suspicious behavior exhibited by these highly-suspected bot accounts with respect to account profile information, friends, followers, interaction with other accounts, tweet content, and posting behavior. We also familiarized ourselves with bot characteristics and behaviors reported in previous studies \cite{varol2017online,abokhodair2015dissecting,chavoshi2017demand,ferrara2016rise,beskow2018bot,stieglitz2017social}. Following this, we have created a list of bot characteristics described in Table~\ref{bot_characteristics}.

Based on this list of bot criteria (Table \ref{bot_characteristics}), we manually examined each account and assigned a bot-likelihood score (we refer to this as the true score) ranging from 0 to 5, with 0 being ``very unlikely" and 5 being ``very likely" based on the extent by which an account exhibited a suspicious bot-like behavior from the list. We also added to the list other suspicious behaviors that we encountered while studying and labeling accounts in our dataset. It is important to note that even human accounts do exhibit one or more of these characteristics at different times (e.g., having a large number of followers). Furthermore, a bot may exhibit only a subset of these characteristics in addition to some human-like characteristics. Therefore, in our manual labeling of bot-like scores, the more characteristics an account exhibited, the higher the bot score it got assigned.

Since manual labeling is time and effort consuming, we considered a sample subset of accounts with hateful tweets. Using a 95\% confidence level and a 4\% margin of error, a representative sample of these accounts would be of size 450. To eliminate sampling bias and to ensure that the sample preserves the statistical proportions of the original dataset (see Figure \ref{Botometer_scores}), we applied proportionate stratified random sampling, wherein simple random sampling technique is employed to select training examples proportionally from each stratum (i.e., subgroup). This sampling method ensures that accounts with unusually high Botometer scores are still present in our sample and in a proportion similar to that in our original dataset. The final sample consisted of 239 accounts from the 0-1 stratum, 95 accounts from the 1-2 stratum, 47 accounts from the 2-3 stratum, 32 accounts form the 3-4 stratum, and 37 accounts from the 4-5 stratum.

Finally, to validate the robustness of our labeling process, we calculated the inter-rater agreement score between the two labelers
on a subset of 30  independently-labeled accounts.
A weighted kappa \cite{cohen1968weighted} score of 0.86 was reported, which indicates an almost perfect agreement \cite{landis1977measurement}. Given such a high inter-rater agreement score, a well-defined bot criteria (Table \ref{bot_characteristics}), and a highly time-expensive task (each account required on average a 15-min examination before a score was given), we decided to split the 450 accounts equally between the two labelers. 


\begin{table}
  \caption{Characteristics and behavior commonly exhibited by bots.}
  \label{bot_characteristics}
  \begin{tabularx}{\linewidth}{L}
    \toprule
     \textbf{Tweeting behavior}\\
     \midrule
     Never replying to or interacting with other accounts. \\
     Retweeting and liking every tweet in a specific hashtag. \\
     Tweeting/retweeting in large volumes. \\
     Tweeting regularly on specific days of the week. \\
     Tweeting regularly at specific times (in hours and minutes) of the day. \\
     Tweeting extensively in the same hashtag. \\
     Tweeting the same content at different times. \\
    Exclusively liking tweets posted by certain accounts. \\
     Liking their own tweets. \\
     Getting few or no likes and/or retweets while having a large number of followers. \\
     Their tweets receive the exact number of retweets, likes, and/or replies from the same accounts.\\

     \toprule
     \textbf{Content and topic characteristics}\\
     \midrule
      Tweeting nonsensical tweets or random letters that do not form actual words (Figure \ref{BotsScreenShots:a} and \ref{BotsScreenShots:c}). \\
     Their tweets appear to be focused on a single specific topic.\\
     Using more than one hashtag for every tweet (Figure \ref{BotsScreenShots:a}). \\
  
     Repeatedly posting links to news articles, books, or videos from a particular source.\\
     Most of their tweets contain pictures and/or URL links. \\
     The same content is posted by several other accounts around the same time. The content usually is about general facts or famous quotes (Figure \ref{BotsScreenShots:b}).\\
    \toprule
     \textbf{Account profile characteristics}\\
     \midrule
     Followed by accounts with an empty profile description (Figure \ref{BotsScreenShots:d}).\\
     They have a very few or a very large number of followers. \\
     Following accounts randomly with no pattern on the kind of accounts they are interested in. \\
     Following zero accounts. \\
     Following accounts who haven't tweeted anything.  \\
     Their screen name doesn't match their Twitter handle.  \\
     They have a long Twitter handle with a scramble of letters and numbers.\\
     Recently joining twitter. \\
     The time zone associated with their tweets doesn't match the location specified on their profile. \\
    \bottomrule
    \end{tabularx}
\end{table}

\begin{figure}[ht]
\subfloat[]{\includegraphics[width=0.48\linewidth]{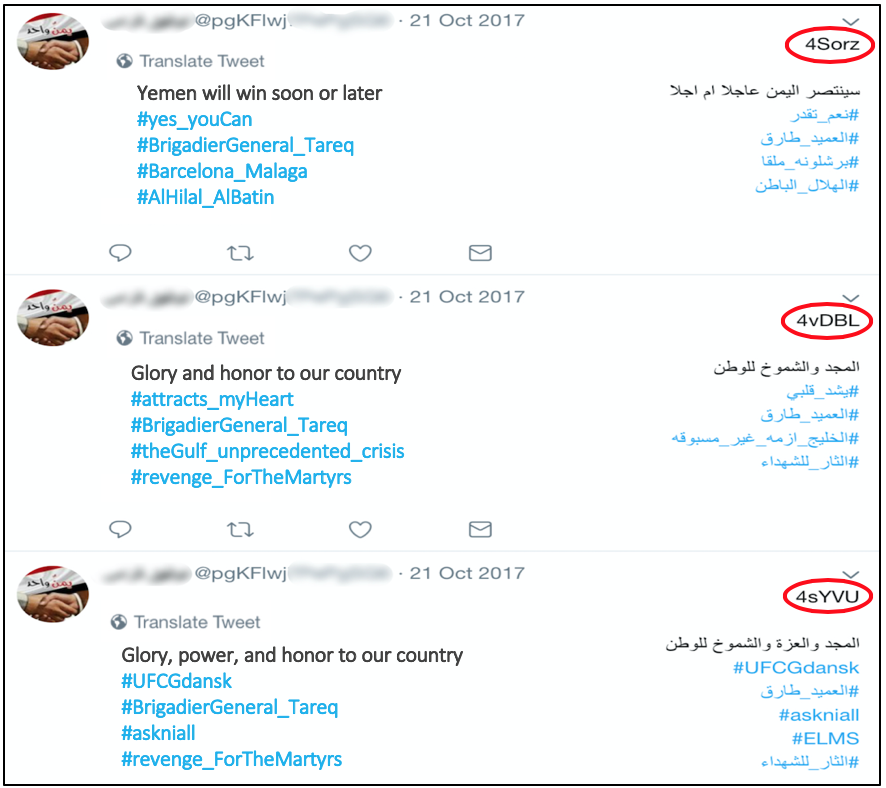}
\label{BotsScreenShots:a}}
    \hfill
\subfloat[]{\includegraphics[width=0.48\linewidth]{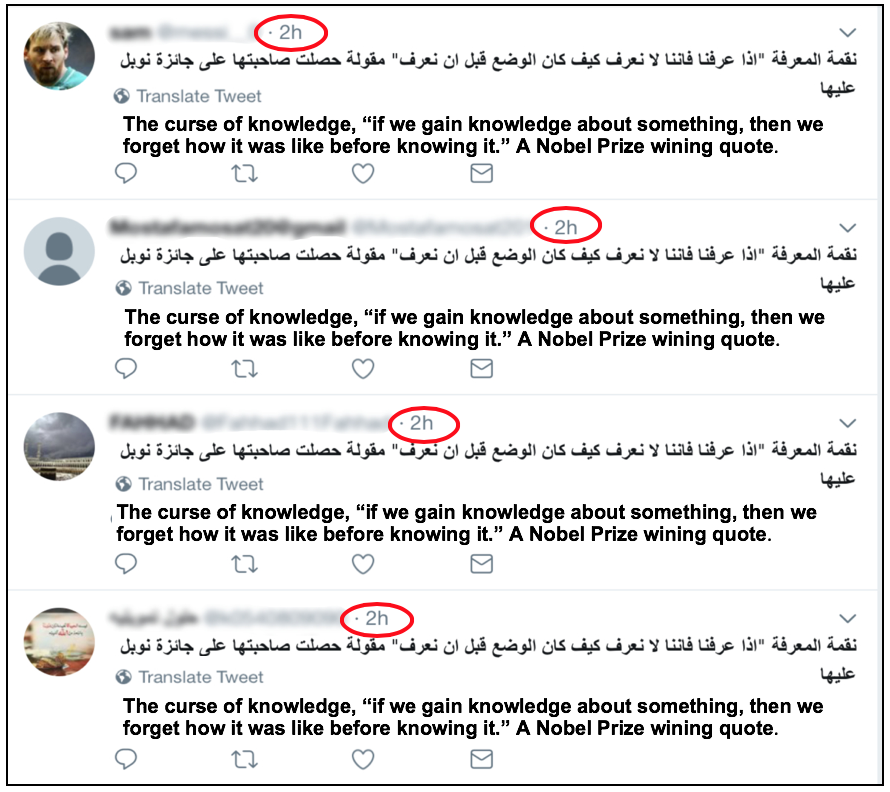}
\label{BotsScreenShots:b}}

\medskip
\subfloat[]{\includegraphics[width=0.48\linewidth]{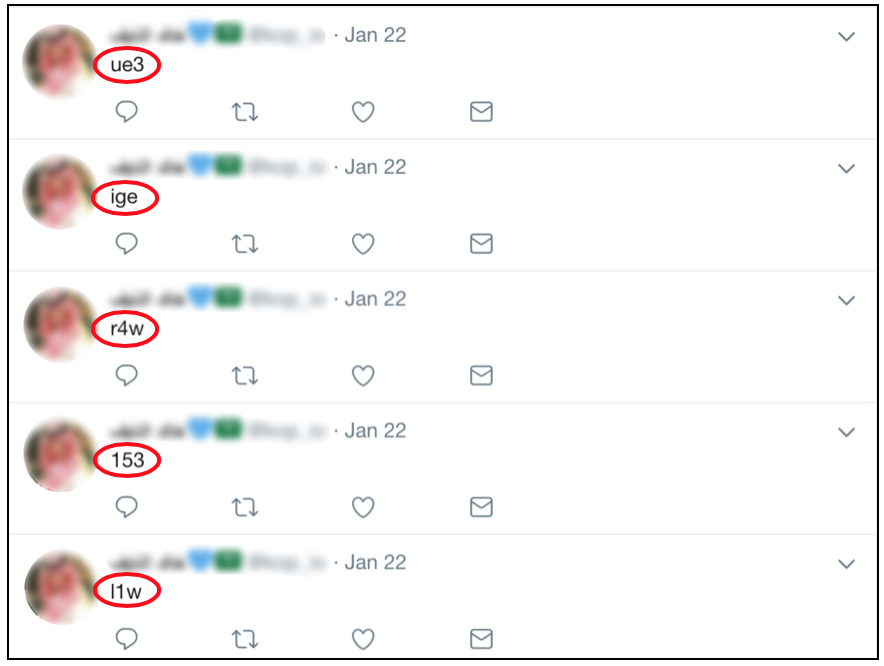}\label{BotsScreenShots:c}}%
    \hfill
\subfloat[]{\includegraphics[width=0.48\linewidth]{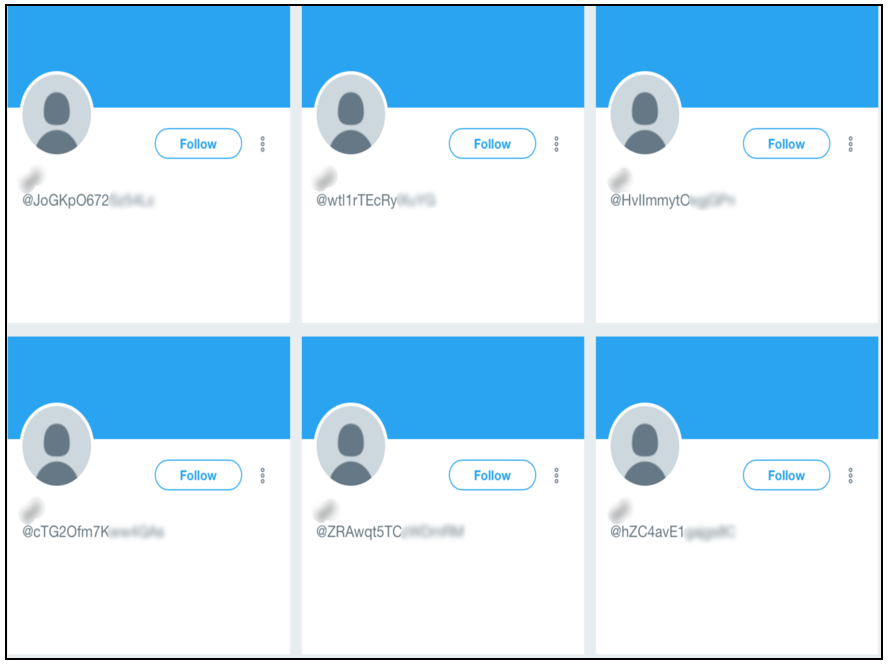}\label{BotsScreenShots:d}}

\caption{Examples of suspicious bot-like behavior.}
\label{BotsScreenShots}
\end{figure}

\subsection{Quantifying Hate Speech Sent by Bots}\label{quantifying}
The results of our manual labeling of the 450 accounts can provide a preliminary indication of how many hateful tweets were sent by bots vs. humans. Assuming that accounts with a true score of 3 or higher were bots, we found that there were 77 (17\%) bots and 373 (83\%) humans. Bots authored 109 hateful tweets (a per-bot average rate of 1.4 tweets), and human accounts authored 446 hateful tweets (a per-human average rate of 1.2 tweets). The ratio of tweets sent by bots to those sent by humans is 1:4. In other words, bots were responsible for 22.6\% of hateful tweets, while humans were responsible for 77.4\% of hateful tweets. 

The relatively low per-bot average rate of tweets could be attributed to the fact that we are only considering their tweets in the hate speech dataset. Considering their whole timeline (tweets) and finding how many of those contain an instance of religious hatred is worth investigating in the future. We will extend this analysis in Section \ref{RoleBotsExtended} to include all 1750 accounts with hateful tweets.

\section{Botometer Assessment}
\subsection{Methods}\label{BotometerAssesment}
Our manual scoring of accounts as well as Botometer scoring is done on a scale of 0-5 with a higher score implying a higher likelihood of the account being a bot. However, the absolute scores assigned by the two scoring methods may differ. In order to evaluate the accuracy of Botometer, we need to investigate if there is a monotonic relationship between how we score accounts (true scores) and how Botometer scores accounts (Botometer scores). To do this, we applied two rank correlation tests that measure the strength and direction of the association between the two scorings.

The first rank correlation test is Spearman's rho ~\cite{spearman1904proof}, which is a well-known nonparametric statistical test that measures the strength of correlation between two independent variables. As Spearman's rho calculation is based on squaring differences between rankings, it penalizes more for large discrepancies between rankings. In case of tied ranks, the mean of the ranks is assigned. The second evaluation metric is Kendall's tau ~\cite{kendall1938new}, which is also a non-parametric test that is used to measure the strength and direction of correlation between two independent sets of scores. The Tau-b version of Kendall's tau was used to handle tied ranks. While both Spearman's rho and Kendall's tau are based on the ranks of data and not the actual scores, Kendall's tau appears to be less sensitive to wide discrepancies between rankings. The value of both Spearman's rho and Kendall's tau ranges from -1 (perfect negative correlation) to +1 (perfect positive correlation). A value of zero indicates the absence of correlation between the two sets of scores. Finally, we applied mean absolute error (MAE) to measure the average absolute differences (errors) between the true scores and Botometer scores. 

\subsection{Results}\label{botometer-analysis}
The results of Spearman's rho and Kendall's tau were 0.43 and 0.33, respectively. The results suggest that there is a moderate positive monotonic relationship between the two sets of scores. MAE was 1.14, which indicates that Botometer scores on average are off by 1.14 points. These results indicate that while Botometer performs better than average in detecting Arabic bots, there is a need for developing social bot detection models that can work more effectively on Arabic Twitter accounts. 

\begin{figure}[ht]
  \centering
  \includegraphics[width=7cm,height=7cm,keepaspectratio]{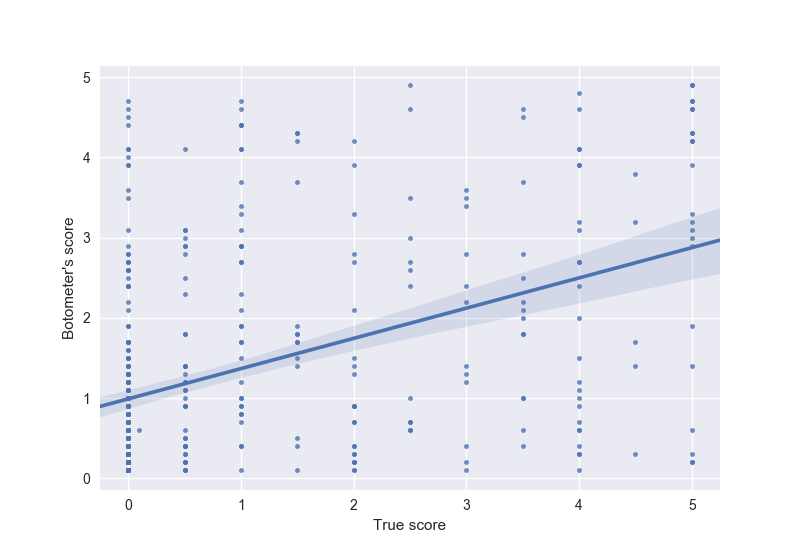}
  \caption{Scatter plot of true score vs. Botometer score, along with a trend line.}
  \label{trendline}
\end{figure}

\begin{figure}[ht]
  \centering
  \includegraphics[width=7cm,height=7cm,keepaspectratio]{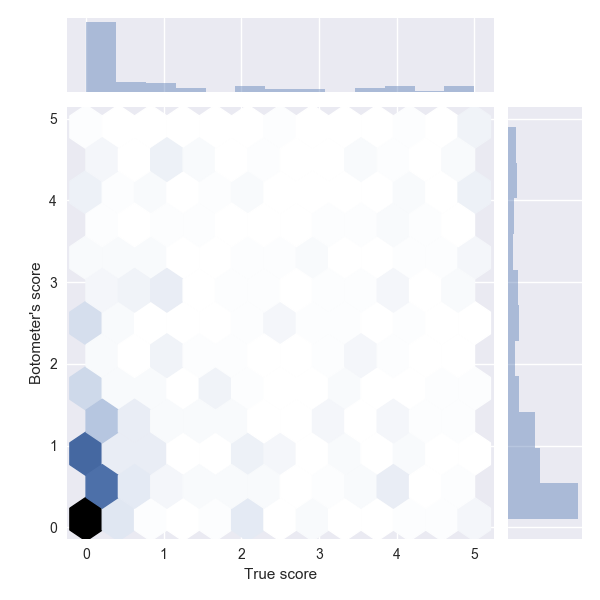}
  \caption{Joint histogram for true score and Botometer score using hexagonal binning.}
  \label{histogram}
\end{figure}

To better understand the limitations of Botometer in detecting Arabic bots, we further analyze the results in Figures \ref{trendline} and \ref{histogram}. Figure \ref{trendline} presents a scatter graph plotting the true score for an account against its Botometer score, along with a regression line that best fits the data. The regression line indicates that Botometer tends to assign a higher score to accounts with true scores of 1.5 or less. On the other hand, Botometer tends to assign lower scores to accounts with a true score higher than 1.5. The margin increases as the true score rises. In other words, Botometer appears to be compressing the range of scores by assigning obvious human accounts a higher score than zero and highly suspected bot accounts a lower score than 5. Figure \ref{histogram} shows a joint histogram for the two sets of scores using hexagonal binning. The figure represents a heatmap where darker colored hexagons represent higher density. As shown in the graph, the highest agreement is when both the true score and Botometer score are between 0 and 1. We can also see from the two histograms that Botometer scores have higher frequencies in the middle bins than the true scores. 

Finally, in order to gain some insights into the reasons for Botometer's weakness in identifying Arabic bots, we manually inspected accounts with wide discrepancies in their true scores and Botometer scores and have identified the following possible reasons for this. Note that while some of these reasons may be applicable to English bots as well, we verified them to some extent only for Arabic bots. It is also important to note that a larger, more structured investigation would be required in order to fully validate those reasons. We leave that as part of our future work.

\begin{enumerate}

    \item \textbf{Botometer appears to be assigning a high-bot score to human accounts who have used their Twitter account for a short period of time, and thus have fewer tweets in their timeline.} This could be due to the restriction Botometer enforced on their data collection process, which is considering only accounts with at least 200 tweets \cite{varol2017online}. We had 17 accounts in our dataset with at most 100 tweets (i.e., inactive), and 71\% of them were given Botometer scores larger than 2.5 while their true scores were less than 2.5. Given that Botometer generally assigned 13\% of all accounts in our dataset scores in the upper range, while their true scores were in the lower range, we found this 71\% misclassification rate to be significant\footnote{We applied N-1 correction for ${\chi}^2$ tests throughout this section.} (${\chi}^2$ = 46.9, df = 1, $p$-value < 0.001).  
    \item \textbf{Having unusually small number of followers or friends (followings) appears to be triggering Botometer to assign a high-bot score without taking into considerations other human-like behavior an account is exhibiting.} We had 29 accounts in our dataset with followers or friends less than 5, 35\% of them were misclassified by Botometer. This was significantly different from what is expected (${\chi}^2$ = 11.4, df = 1, $p$-value < 0.001). 
    \item \textbf{As Botometer doesn't support Arabic language, it may miss linguistic and content cues that can giveaway bots, e.g., posting unrelated tweets to a hashtag.} We show in Section \ref{regressionResults} that linguistic features such as the use of numerics and emojis can be significant distinguishing features. This could be a reason for Botometer assigning a lower score to bot-like accounts with higher true scores.
    \item \textbf{Sometimes Arabic Twitter accounts use third-party Islamic applications that post Islamic supplications and/or Quranic verses on their behalf.} There were 18 unique Islamic applications that were used by accounts in our dataset. Such behavior may result in Botometer assuming that these accounts are bots although some of them are in fact humans. This could be a reason for Botometer assigning a higher score to obvious human accounts with true scores of 1.5 or less.
\end{enumerate}
    We also considered other reasons that we believed to be causing wide discrepancies between true scores and Botometer scores. For example, including a hashtag in every tweet appeared to be triggering Botometer to assign a high bot score even when the account exhibited more human-like behavior. We had 41 accounts in our dataset with an average of one or more hashtags per tweet, and Botometer assigned higher scores to 15\% of them. However, we found this proportion to be statistically insignificant. Another case where we noticed higher scores given to accounts by Botometer is when human accounts appeared to be followed by fake (probably purchased) followers. However, we couldn't verify this claim as such feature (followers being fake or not) was not part of our collected metadata. 
\section{Detecting Arabic Language Bots}
\subsection{Methods}
We now present a regression predictive modeling task in which we train a random forest regression model to predict the bot score, from zero to 5, of an account based on various hand-crafted features. Random forest is a tree-based ensemble method that has been successfully used in other bot detection tasks \cite{davis2016botornot,beskow2018bot}. A great property of random forest is that it inherently computes feature importance, and thus it can give insights into which feature is most informative. Random forest can also control over-fitting by training each tree on randomly selected training examples and features~\cite{Rfoverfitting}. Note that we also experimented with other regression algorithms such as logistic regression and gradient boosting regression trees. However, their performance was poorer compared to random forest, and thus we only report the results achieved by random forest.

We implemented random forest with scikit-learn \cite{scikit-learn}, a Python machine learning library. To understand the impact of each individual feature in detecting bots, we trained our regression model on successive combinations of content, tweet, topic, and account features. We tuned the regression model by performing a hyperparameter grid search with 10-fold cross-validation and optimized for Spearman's rho. 
The regression model was trained on 70\% of the accounts and tested on the remaining 30\%. Three evaluation metrics were used to compare the performance of our regression model to that of Botometer: Spearman's rho, Kendall's tau, and MAE, as discussed in Section~\ref{BotometerAssesment}.

\subsection{Features}\label{features}

\begin{table}[ht]
  \caption{Summary of features used for training the regression model.}
  \label{features_table}
  \begin{tabularx}{\linewidth}{llL}
    \toprule
    \textbf{Category} & \textbf{Feature} & \textbf{Description} \\
    \midrule
    \multirow{9}{*}{Content Features} & avg\_word\_len & The average length of words\\
                                      & avg\_tweet\_len & The average length of tweets\\
                                      & avg\_punctuations & The average number of punctuations per tweet\\
                                      & avg\_emojis & The average number of emojis per tweet\\
                                      & avg\_numerics & The average number of numerical values per tweet\\
                                      & avg\_hashtags & The average number of hashtags per tweet\\
                                      & avg\_mentions & The average number of account mentions per tweet\\
                                      & avg\_links & The average number URL links per tweet\\
                                     & avg\_repetition & The average number of elongated (lengthened) words per tweet (e.g., thaaanks)\\
    \midrule
    \multirow{5}{*}{Tweet Features}  & original\_proportion & The proportion of original tweets\\
                                     & reply\_proportion &  The proportion of reply tweets\\
                                     & retweet\_proportion &  The proportion of retweet tweets\\
                                     & got\_retweeted &  The number of times their tweets got retweeted\\
                                     & got\_ favorited & The number of times their tweets got favorited\\
    \midrule
    \multirow{9}{*}{ \pbox{20cm}{Topic \& Sentiment \\ Features}} & polarity & A sentiment score from -1 to +1 \\
                                    & subjectivity & A score from 0 (objective) to 1 (subjective)\\
                                    & topic\_1 & Topic 1 (Sports) distribution\\
                                    & topic\_2 & Topic 2 (General/Videos) distribution\\
                                    & topic\_3 & Topic 3 (Political/Houthi) distribution\\
                                    & topic\_4 & Topic 4 (General/Books) distribution\\
                                    & topic\_5 & Topic 5 (Supplications) distribution\\
                                    & topic\_6 & Topic 6 (Political/Jew) distribution\\
                                    & topic\_7 & Topic 7 (Political/Jerusalem) distribution\\
    \midrule
    \multirow{12}{*}{account Features} & statuses\_count &  The number of all tweets on the account (acct)\\
                                    & followers\_count & The number of accts following the acct \\
                                    & friends\_count & The number of accts the acct is following\\
                                    & favorite\_count & The number of tweets favorited by the acct\\
                                    & account\_age & The acct's age in months\\
                                    & empty\_description? & Does the acct have an empty profile description?\\
                                    & empty\_location? & Does the acct have an empty location?\\
                                    & empty\_URL? &  Does the acct have an empty link in their profile?\\
                                    & default\_profile? &  Does the acct have the default profile background?\\
                                    & default\_image? &  Does the acct have the default profile image?\\
                                    & verified? & Has the acct's identity been verified by Twitter?\\
                                    &geo\_enabled? & Does the acct have the geotagging feature enabled?\\
  \bottomrule
\end{tabularx}
\end{table}

Based on our analysis in Section ~\ref{botometer-analysis}, we identify four sets of features that we anticipate to be informative for Arabic bot detection. These are content, tweet, topic \& sentiment, and account features. Table \ref{features_table} provides a description of each of these features. For content features, we used average lengths of words and tweets, and average numbers of emojis, punctuation marks, numerics, hashtags, account mentions, URL links and elongated words per tweet. For tweet features, we used the proportions of original tweets, reply tweets and retweet tweets, as well as the number of times an original/reply tweet was retweeted or favorited. For account features, we considered features such as the total number of tweets posted and favorited, numbers of followers and friends as well as features related to the account profile (e.g., account age).

Sentiment features were obtained by using TextBlob NLP Python library \cite{loria2018textblob} which offers support for the Arabic language. Topic modeling was implemented using Latent Dirichlet Allocation (LDA) model provided by Gensim \cite{rehurek_lrec}, an unsupervised topic modeling python library. Before extracting topic features, tweets were preprocessed by removing diacritics (i.e., accents), tatweel (i.e., kashida), elongation, two-character words, non-Arabic characters, URL links, punctuation marks and numbers. We also removed Arabic stop words and normalized Arabic letters and hashtags as described in \cite{albadi2018they}, and then filtered out very rare words that appeared in less than 10 accounts' timelines and too common words that appeared in more than 50\% of accounts' timelines. From the remaining list, we considered the 10K most frequent words. We experimented with both bag-of-words (bow) and term frequency-inverse document frequency (tf-idf) text representation techniques. We also experimented with stemming words using ARLSTem Arabic stemmer from NLTK Python library~\cite{bird2009natural} and lemmatization using StanfordNLP Python library~\cite{qi2018universal}. Results of these experiments are provided in the next subsection.

\subsection{Results}\label{regressionResults}
We trained regression models on successive sets of features and assessed their generalization ability on a holdout testing dataset. Although we collected up to 3,200 tweets for each account, training the regression model using up to 200 tweets from each account provided faster training with similar results. Therefore, the results reported here are the ones using features extracted from up to 200 tweets per account, resulting in a total of 86,346 tweets.  

Table \ref{results} compares the performances of these regression models in terms of Spearman's rho, Kendall's tau, and MAE. Highest scores are shown in \textbf{bold}. We have included the performance of Botometer as well in this table as a baseline. As shown in the table, our regression
model trained on only simple content features outperformed Botometer which uses user, friend, network, and temporal features. The most informative content features reported by this regression model were the average numbers of account mentions, URL links, numerics, and punctuation marks, respectively. This shows that linguistic cues conveyed in tweets are highly effective in detecting Arabic bots. We will further discuss the importance and direction of contribution of these features in Section \ref{FeatureImportance}.

By including the tweet features in addition to the content features in training the regression model, the Spearman's coefficient improved by five points. Among the content and tweet features, the most distinguishing features were reply tweet proportion, original tweet proportion, and average number of account mentions, respectively. Adding topic and sentiment features in training further improved the performance of our regression model. These topic features were extracted using bow as opposed to tf-idf, as bow delivered better performance. We found that topic features extracted from lemmatized text achieved superior results to those extracted from stemmed text. However, we also found that not using stemming or lemmatization led to the best performance. 

The best Spearman's rho and Kendall's tau were achieved after adding account features. The 0.74 in Spearman's rho indicates a strong positive correlation between scores predicted by our regression model and the true scores. The most informative features for this regression model were still reply tweet proportion, average number of mentions, and original tweet proportion, respectively. 

The least informative features were mostly from the account feature category such as whether the account has an empty profile description, location, and URL link. Also, whether or not the account has the default profile image or their geotagging feature enabled didn't contribute much to the predicted bot score. This suggests that there wasn't a significant difference between the distribution of humans and bots across those features. 

\begin{table}[ht]
  \caption{Evaluation results of various regression models.}
  \label{results}
  \begin{tabular}{llll}
    \toprule
    \textbf{Model/Features} & \textbf{Spearman's rho} & \textbf{Kendall's tau} & \textbf{MAE}\\
    \midrule
    Botometer & 0.43 & 0.33 & 1.14\\
    Content Features & 0.63 & 0.50 & 0.85\\
    + Tweet Features & 0.68 & 0.55& 0.72\\
    + Topic \& Sentiment Features & 0.72 & 0.59& \textbf{0.69}\\
    + Account Features & \textbf{0.74} & \textbf{0.60} & 0.75 \\
    \bottomrule
    \end{tabular}
\end{table}

\section{Analysis}
\subsection{Feature Importance \& Contribution}\label{FeatureImportance}
Random forest computes a feature importance score for each feature using the mean decrease impurity feature selection method which reflects how much each feature reduces variance \cite{impurity}. The top most-important features, along with their importance scores for the best performing regression model are shown in Table \ref{important_features}. 

\begin{table}[ht]
  \caption{Most informative features for distinguishing bots.}
  \label{important_features}
  \begin{tabular}{llll}
    \toprule
    \textbf{Feature} & \textbf{Importance} & \textbf{Feature} & \textbf{Importance}\\
    \midrule
    1. reply\_proportion & 0.136 & 6. favorite\_count & 0.047\\
    2. avg\_mentions & 0.104 & 7. avg\_links  & 0.046\\
    3. original\_proportion & 0.080 & 8. avg\_numerics & 0.042 \\
    4. retweet\_proportion & 0.066 & 9. friends\_count & 0.034 \\
    5. avg\_emojis & 0.049 &  10. avg\_repetition & 0.033\\
    \bottomrule
    \end{tabular}
\end{table}

\begin{figure}[ht]
\subfloat[Relpy tweet proportion]{\includegraphics[width=0.33\linewidth]{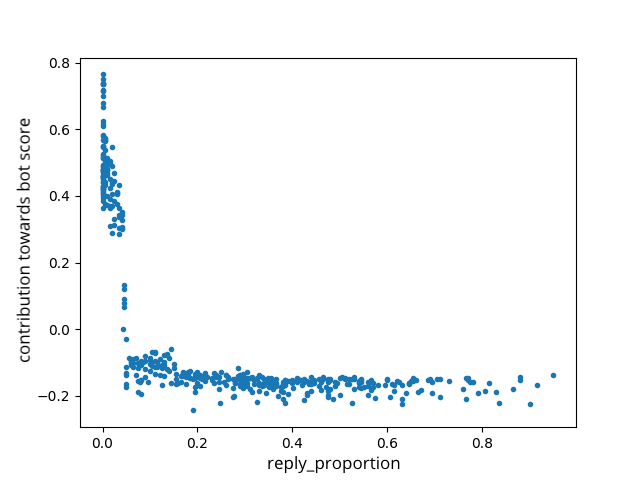}\label{Tree:a}}
    \hfill
\subfloat[Average number of account mentions]{\includegraphics[width=0.33\linewidth]{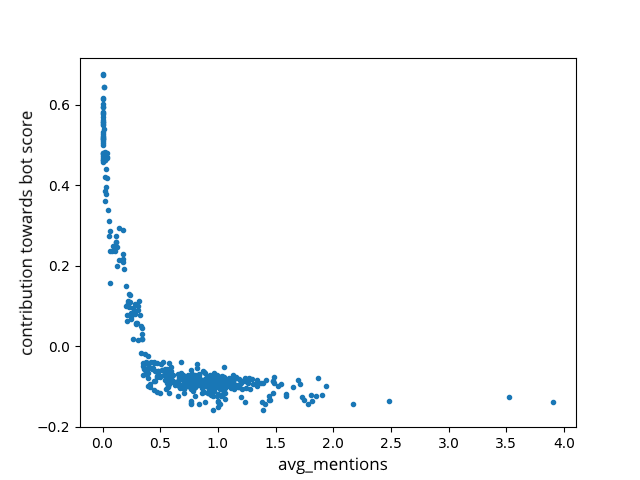}\label{Tree:b}}
\subfloat[Original tweet proportion]{\includegraphics[width=0.33\linewidth]{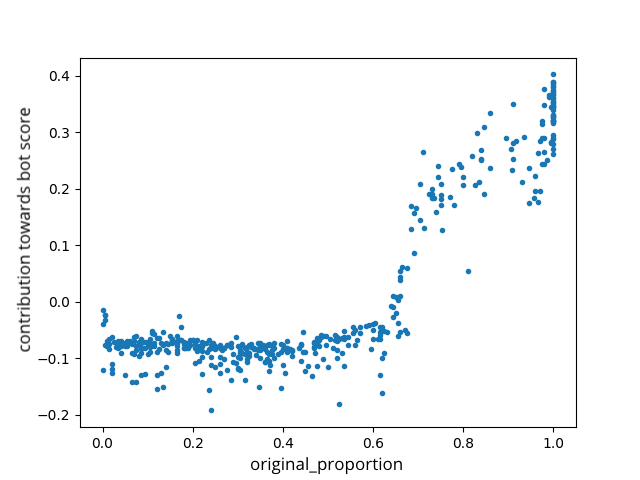}\label{Tree:c}}
\medskip

\subfloat[Retweet tweet proportion]{\includegraphics[width=0.33\linewidth]{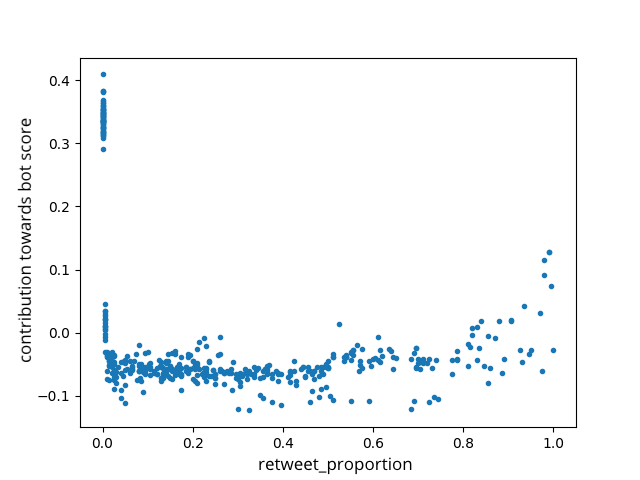}\label{Tree:d}}
    \hfill
\subfloat[Average number of emojis]{\includegraphics[width=0.33\linewidth]{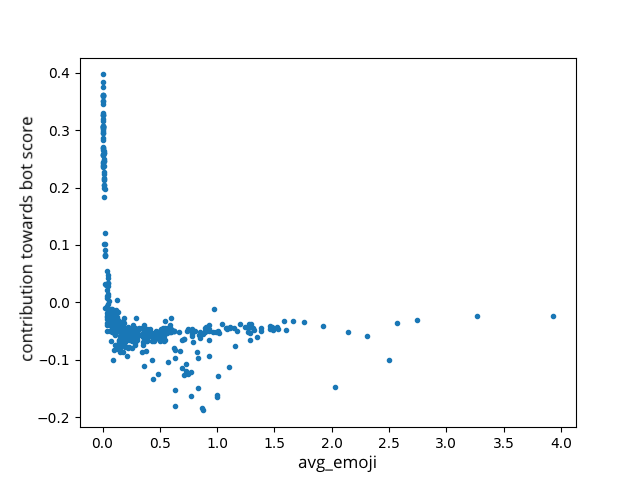}\label{Tree:e}}%
\subfloat[Average number of URL links]{\includegraphics[width=0.33\linewidth]{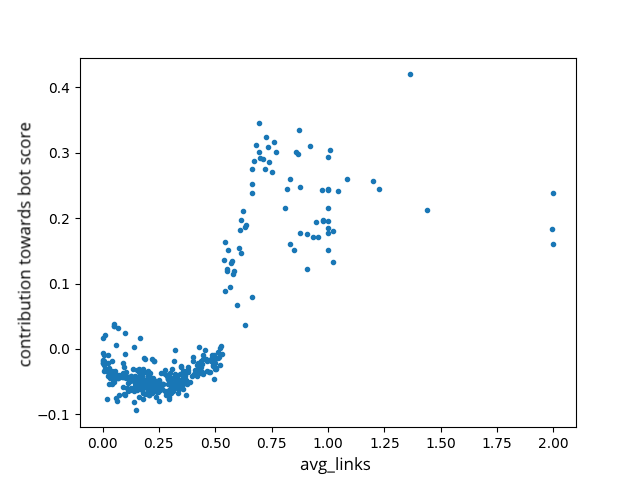}\label{Tree:f}}
\medskip

\subfloat[Average number of numerics]{\includegraphics[width=0.33\linewidth]{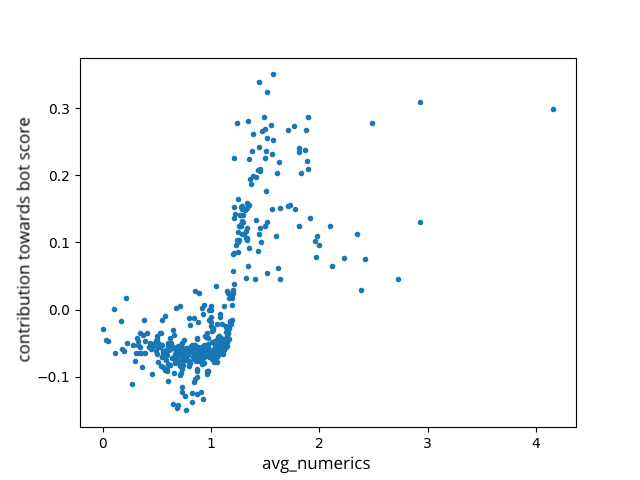}\label{Tree:g}}
\hfill

\caption{Feature contribution for some of the most informative features.}
\label{TreeinterpretGraphs}
\end{figure}

Random forest feature importance score doesn't offer insights into feature contribution (i.e., the direction of feature importance). To understand how much positively or negatively each feature contributed to the final predicted bot score, we used TreeInterpreter Python package \cite{TreeInterpreter}, which breaks down each prediction made by the regression model into the sum of bias (i.e., the average bot score in the training data) and each features' contribution. We selected some of the top informative features and plotted their contribution against their corresponding feature value (see Figure \ref{TreeinterpretGraphs}). Figure \ref{Tree:a} shows feature contribution for the reply tweet proportion which is the most bot-distinguishing feature. It shows that the more reply tweets an account has, the less likely that account is a bot. This suggests that these bots are not yet smart enough to engage in conversations and interact with other accounts as humans would normally do. The same feature contribution pattern was found for the average number of mentions as illustrated in Figure \ref{Tree:b}. Mentioning other accounts usually implies interacting and communicating with them, and thus the more social an account is, the less likely that account is a bot. 

Figure \ref{Tree:c} shows how the proportion of original tweets (not retweet or reply tweets) contributes to the predicted bot score. If original tweets constitute more than 60\% of an account's overall tweets, then the predicted bot-likelihood score would increase as the original tweet percentage increases. Such accounts that don't reply nor retweet might be using third-party applications to post tweets on their behalf, or that their ``masters'' programmed them so that they only disseminate prespecified text. This also suggests that human accounts on Twitter usually exhibit a variety of behaviors such as replying, retweeting, and tweeting an original text. As for retweeting, we can see from Figure \ref{Tree:d} that there are two retweeting behaviors that would result in an increase in the predicted bot score. These are never retweeting any tweet (x $\approx$ 0) and retweeting extensively (x $\approx$ 1). Again, such black and white behavior is more of a bot-like behavior rather than a human-like behavior. 

We also found a clear distinction between English bots and Arabic bots in terms of retweet, reply, and original tweet proportions. It has been claimed that English bots tend to retweet more than posting original tweets \cite{bessi2016social}. This was not found to be true in our dataset, i.e., Arabic bots in our dataset were posting original tweets more often than retweeting tweets. In particular, the average retweet, original, and reply proportions for bots were 17\%, 76\%, and 7\%, respectively.  

The average number of emojis per tweet was also one of the highly informative features. This feature was not considered by Botometer because it wasn't trained for Arabic bots. Figure \ref{Tree:e} illustrates that bots tend to not use emojis in their tweets. We believe that this could be due to the fact that people use emojis instinctively to convey different kinds of emotions. Having a URL link in more than 50\% of an account's tweets would contribute positively to the predicted bot score as shown in Figure~\ref{Tree:f}. This makes sense as many automatically generated tweets contain links to books, news articles, posts from a linked Facebook account, etc. 

Another feature that might not be considered by Botometer as it doesn't extract Arabic-specific features is the average number of numerics. At first, we were surprised to find that the more numbers accounts use in their tweets, the more likely they are bots (see Figure \ref{Tree:g}). Upon closer look at accounts with a high use of numbers in their tweets and a positive average number of numerics feature contribution, we found that some accounts had random English letters and numbers in their tweets, which suggests that such tweets were generated automatically by computers. Such behavior was previously shown in Figures \ref{BotsScreenShots:a} and \ref{BotsScreenShots:c}.

\subsection{Topic, Source, and Network Analysis}

\subsubsection{Topic Analysis}\label{topicAnalysis}
Here we further investigate the topics that are most discussed by bots (true scores $>$ 2.5) and humans (true scores $\leq$ 2.5). Although LDA doesn't assign names to topics, we inferred those names from the list of terms that are most associated with each topic. To give more insight into what these topics represent, we list in Table \ref{LDAtopics} the most relevant terms for each topic.

\begin{table}[ht]
  \caption{Most relevant terms for each LDA-identified topic.}
  \label{LDAtopics}
  \begin{tabular}{ll}
    \toprule
    \textbf{Topic} & \textbf{Term}  \\
    \midrule
    \multirow{5}{*}{Topic 1 (Sports)} & Al-Hilal FC \\
                                      &  Al-Ahli FC  \\
                                      &  Al-Nassr FC \\
                                      &  Al-Ittihad FC \\
                                      &  football club\\
    
    \midrule
    \multirow{5}{*}{Topic 2 (General/Videos)}  &  video\\
                                               &  I liked\\
                                               & Iraq \\
                                               &  message \\
                                               & the truth\\
    
    \midrule
    \multirow{5}{*}{Topic 3 (Political/Houthi)}  &  Houthi\\
                                                &  Yemen \\
                                                &  Qatar\\
                                                & UAE \\
                                               &  The State\\
    
    \midrule
    \multirow{5}{*}{Topic 4 (General/Books)} &    book \\
                                             &  design \\
                                             &   peace \\
                                             &  episode \\
                                             &  link \\
     \midrule
    \multirow{5}{*}{Topic 5 (Islamic Supplications)}  & glory [be to]\\
                                                     & lord \\
                                                     & peace [be upon]\\
                                                     & prophet\\
                                                     & blessings [be upon]\\
    \midrule
    \multirow{4}{*}{Topic 6 (Political/Jew)}   & the Jews\\
                                              &  death\\
                                               &  Saleh\\
                                               &  Iraq\\
                                              
      \midrule
    \multirow{5}{*}{Topic 7 (Political/Jerusalem)}  &  Jerusalem\\
                                                    &  Palestine \\
                                                    & Syria\\
                                                    & Israel\\
                                                    &  Qatar\\
  \bottomrule
\end{tabular}
\end{table}



\begin{figure}[ht]
  \centering
  \includegraphics[width=\linewidth]{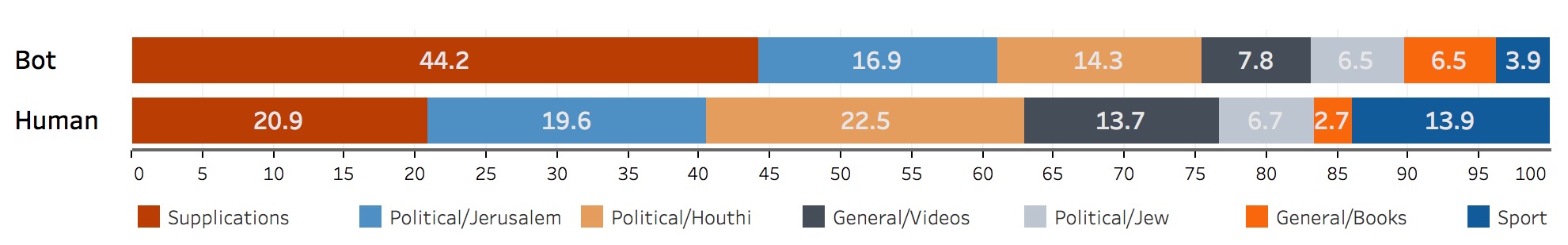}
  \caption{Dominant topic proportions of human vs. bot tweets.}
  \label{dominant_topics}
\end{figure}

 We then considered the most dominant topic for each account, i.e., the topic with the highest probability distribution. Figure \ref{dominant_topics} shows the percentages of dominant topics of humans and bots tweets. We found that the distributions of humans and bots differ significantly among the seven topics (${\chi}^2$ = 25.9, df = 6, $p$-value < 0.001). The topic distributions for bots is lopsided, i.e., the majority of the posts from bots were concentrated on a small number of topics, while humans' tweets covered a wider range of topics. While the top three discussed topics for both humans and bots were identical, the percentages were different. About 44\% of suspected bots were mainly posting Islamic supplications and prayers, while 21\% of humans were tweeting about the same topic. Suspected bots were least interested in sports (3.9\%); however, they showed somewhat similar interest to that of humans in political topics related to Jerusalem, Jews, and Houthi.




\subsubsection{Source Analysis}\label{sourceAnalysis}
Twitter provides source label along with tweet metadata which indicates which source (i.e., client) was used to post the tweet to its service. The accounts in our dataset were posting tweets using various official and/or third-party sources. We considered the dominant (i.e., main) used source for each account and grouped these sources into three categories: official Twitter sources, Islamic supplications, and other third-party sources. Official Twitter sources include Twitter Web Client, Twitter Lite, Twitter for IPhone, Twitter for Android, and Twitter for Windows. Islamic supplications include third-party applications mainly for automatically posting Islamic supplications on accounts' behalf. Other third-party sources include Facebook, Instagram, Google, If This Then That\footnote{\url{https://ifttt.com}} (IFTTT), Tweetbot, and Alameednews.com. In total, bots were mainly tweeting using  17 unique sources, while humans were tweeting using 14 unique sources. 

\begin{figure}[ht]
  \centering
  \includegraphics[width=\linewidth]{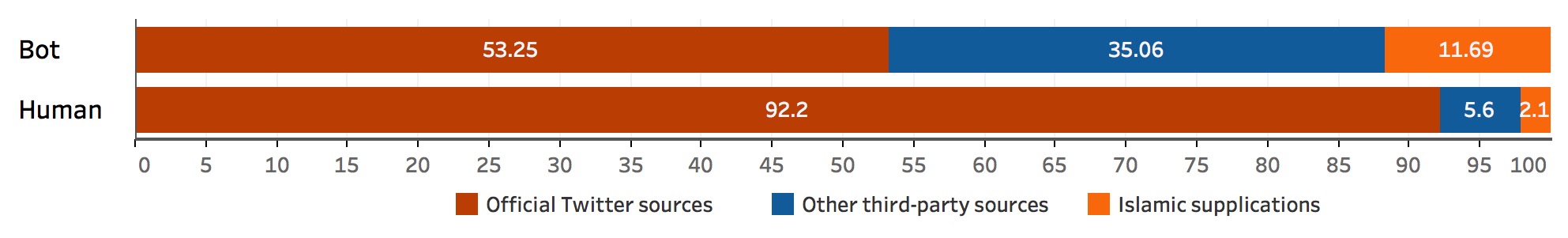}
  \caption{Dominant source proportions for human vs. bot tweets.}
  \label{dominant_sources}
\end{figure}

Figure \ref{dominant_sources} illustrates the percentages of dominant sources for bots and humans accounts. We found that the distributions of humans and bots differ significantly among the three categories of sources (${\chi}^2$ = 78.6, df = 2, $p$-value < 0.001). About 92\% of humans were mainly posting tweets using official Twitter sources, whereas 53\% of bots were mostly using official Twitter sources to post tweets. Posting mainly using third-party sources (including Islamic ones) was a more common behavior of bots (47\%) rather than humans (8\%). 

\subsubsection{Network Analysis}\label{networkAnalysis}
\begin{figure}[ht]
\subfloat[Human]{\includegraphics[width=0.35\linewidth]{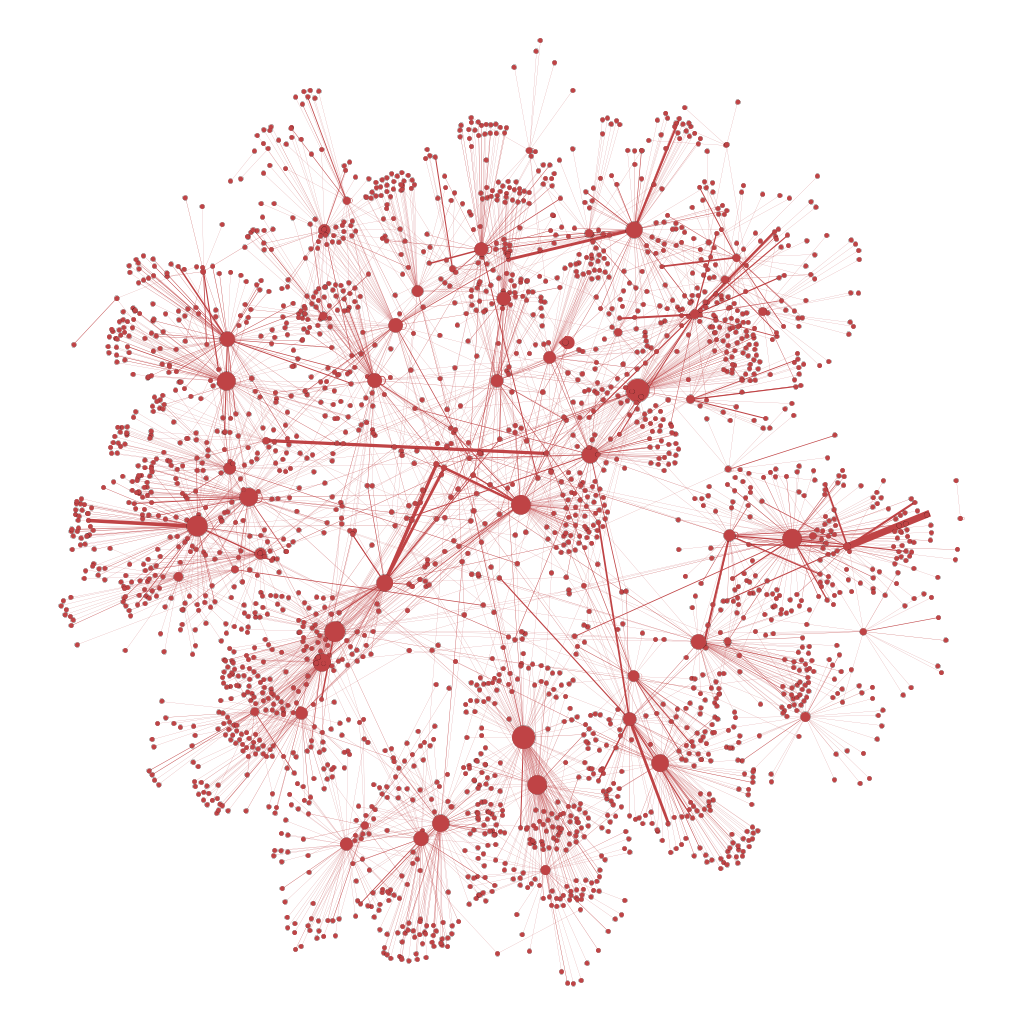}
\label{retweet:human}}
\subfloat[Bot]{\includegraphics[width=0.35\linewidth]{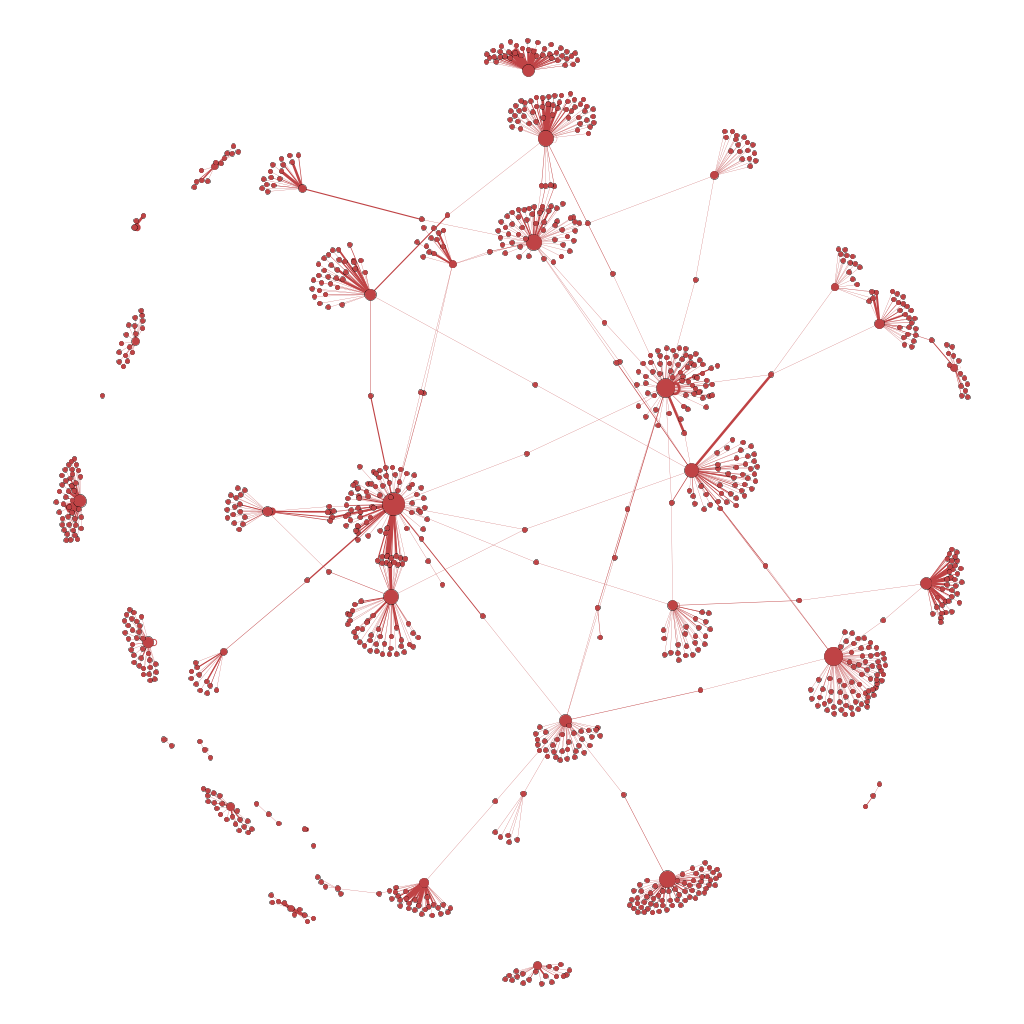}
\label{retweet:bots}}
\caption{Retweet network for human vs. bot.}
\label{retweetNetwork}
\end{figure}

\begin{figure}[ht]
\subfloat[Human]{\includegraphics[width=0.35\linewidth]{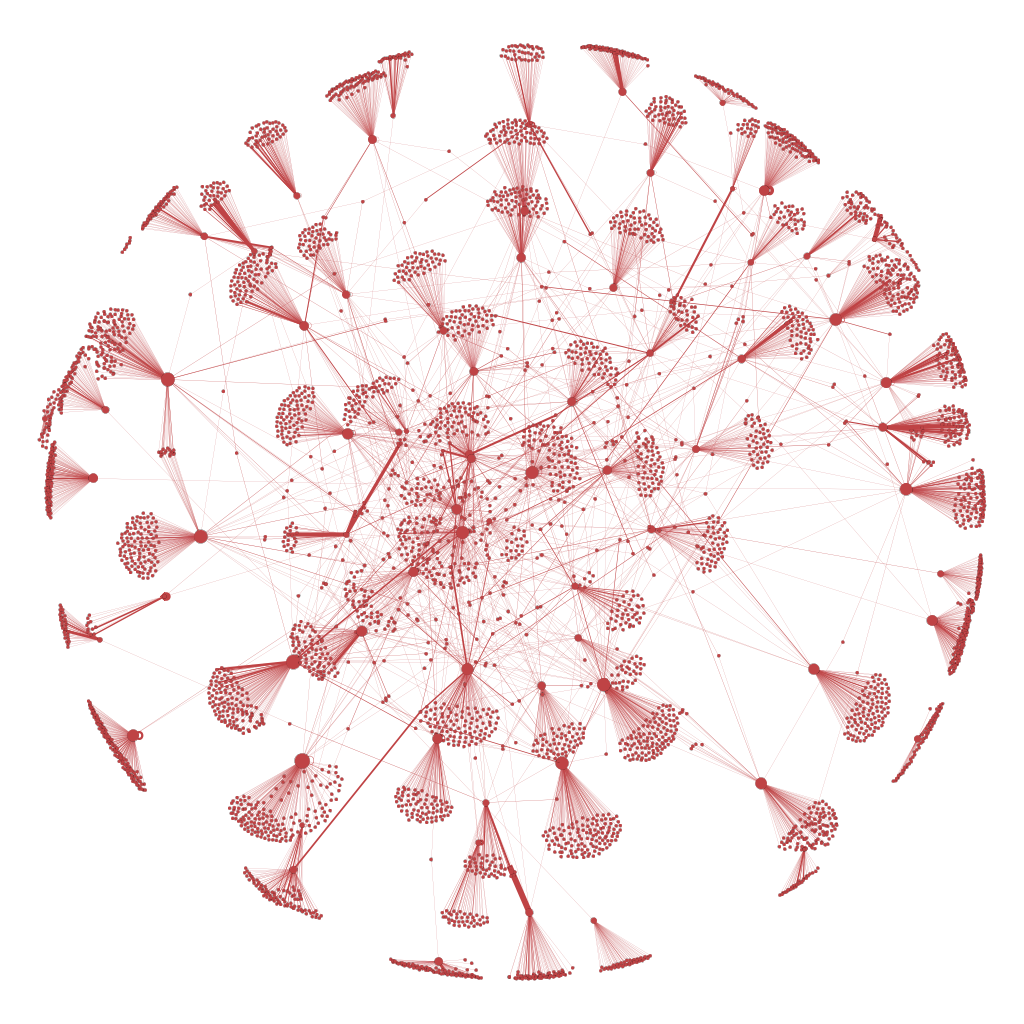}
\label{mention:human}}
\subfloat[Bot]{\includegraphics[width=0.35\linewidth]{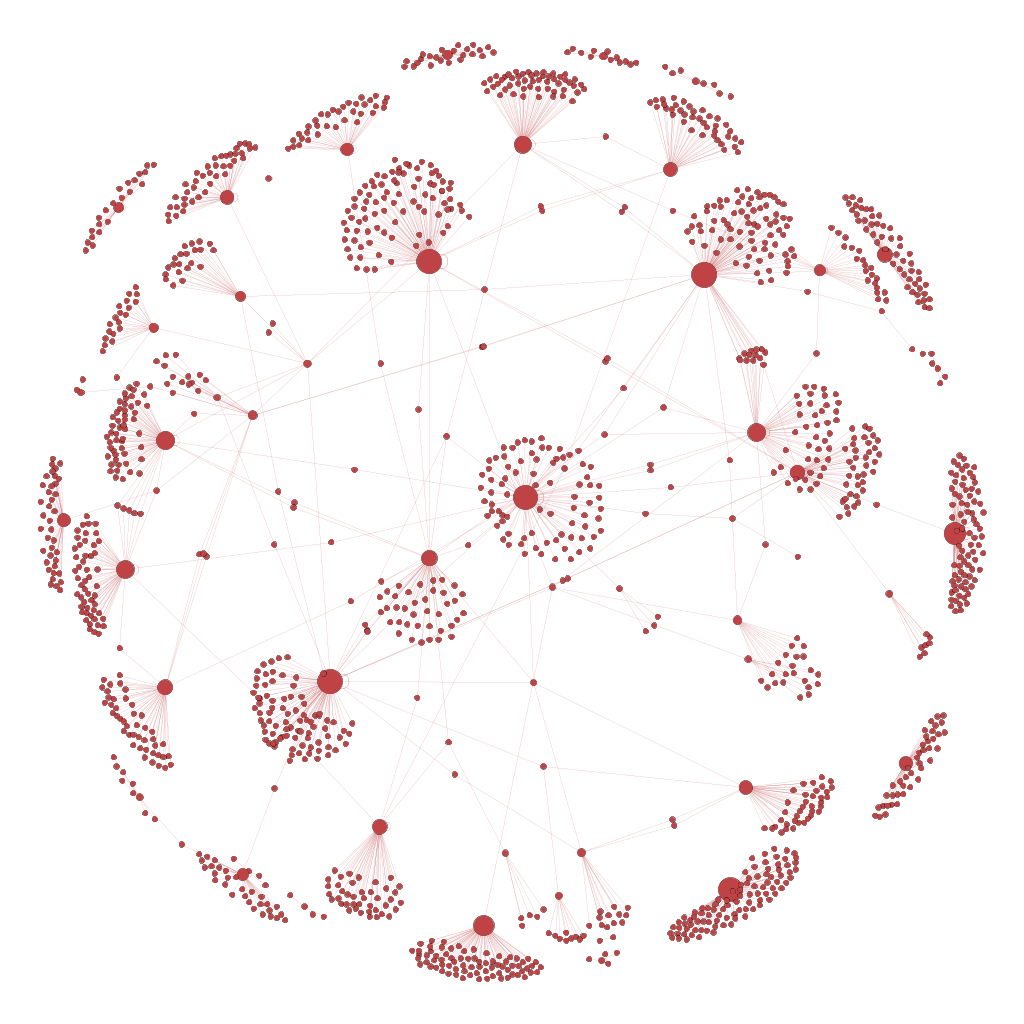}
\label{mention:bots}}
\caption{Mention network for human vs. bot.}
\label{mentionNetwork}
\end{figure}

It has been shown that bot network characteristics differ significantly from those of humans \cite{beskow2018bot}. Here we investigate if this holds for Arabic bots as well. Since in our dataset we have more human accounts (373) than bot accounts (77) and to ensure a fair comparison, we randomly selected 77 human accounts to match the set of the 77 bot accounts. We constructed two types of networks: retweet network (see Figure \ref{retweetNetwork}) and mention network (see Figure \ref{mentionNetwork}). Nodes in these graphs represent accounts in our dataset as well as all other accounts that got retweeted/mentioned by the accounts in our dataset. Edges represent a retweeting/mentioning activity. 

In the retweet network, humans have 2,561 nodes and 2,831 edges, while bots have 1,018 nodes and 1,054 edges, i.e., the human retweet network was more than twice as large as the bot retweet network. This gap was even larger for the mention network; the human mention network (4,978 nodes and 6,514 edges) was more than three times as large as the bot mention network (1,585 nodes 1,666 edges). We can see that bot networks are loosely connected with many singleton nodes, while human networks are highly connected with a very few singleton nodes. These network results are in line with what has been found for English bot networks \cite{beskow2018bot}.

\subsection{The Role of Bots in Spreading Hate Speech}\label{RoleBotsExtended}
To answer the question on how many hateful tweets were sent by bots rather than humans, we extend our analysis presented in Section \ref{quantifying} to include all 1750 accounts with hateful tweets. Three accounts were suspended before we were able to collect their data, and thus they were excluded from this analysis. As we already have true scores for 450 accounts, we applied our regression model to the remaining 1,297 accounts with hateful tweets. Of the 1747 accounts with hateful tweets, we found that 185 (10.6\%) accounts were more likely to be bots (predicted/true scores $>$ 2.5), and 1,562 (89.4\%) accounts were more likely to be humans (predicted/true scores $\leq$ 2.5). Bots authored 238 hateful tweets (1.29 per-bot average rate), whereas humans authored 1,974 tweets (1.26 per-human average rate). The ratio of hateful tweets sent by bots to those sent by humans is 1:8. In particular, humans were responsible for 89.24\% of all hateful tweets, while bots were responsible for 10.76\% of all hateful tweets.  

At the time of this writing (March 2019), we checked to see if the bots identified in our study were still active or not. We found that only 11\% of them were suspended by Twitter. This indicates that the remaining 89\% of the bots have lived for at least 1.4 years. In a recent study by Chavoshi et al. \cite{chavoshi2017demand}, Twitter suspended 45\% of the bots detected in their study within a three-month period. This shows that Arabic bots can go undetected for a long period of time. 

\section{DISCUSSION}\label{Discussion}

Our analysis suggests that Arabic Twitter bots do have a role in the spread of religious hate on Arabic Twitter. In particular, bots were responsible for about 11\% of all hateful tweets in the hate speech dataset. Our topic analysis showed that bots participate in highly controversial political discussions related to Israel/Palestine and Yemen. This builds on prior work that showed participation of Arabic bots, especially through the dissemination of highly polarizing tweets, during the Syrian civil war \cite{abokhodair2015dissecting}.

Such political use of bots (i.e., disseminating hate speech and highly biased news) has been shown to be true for English bots as well. Bots on English Twitter have been used to promote jihadist propaganda \cite{benigni2017online, berger2015isis}, spread fake news \cite{ratkiewicz2011truthy}, and infiltrate political discussions \cite{bessi2016social}. Bots have also been used for spamming in both Arabic \cite{el2016detecting} and English \cite{stringhini2010detecting} Twitter networks. Other nefarious roles of bots that have been explored on English Twitter include manipulating the stock market and stealing personal data \cite{ferrara2016rise}. Unfortunately, there is a significant lack of Arabic-focused research to investigate other roles that can be played by bots in Arabic social networks. Our study serves as a starting point for understanding and detecting Arabic bots, demanding additional research to explore this understudied area.

While the social roles played by Arabic and English bots can be to some extent similar, our analysis showed that some Arabic bot characteristics are unique and different from English bots. As discussed in Section ~\ref{FeatureImportance}, Arabic bots in our dataset were posting original tweets more often than retweeting tweets. This found to be in contrast to English bots that tend to retweet more than posting original tweets ~\cite{bessi2016social}. We also showed that Arabic bots can live longer than English bots. Further, it has been shown that English bots tend to have fewer followers than humans \cite{stieglitz2017social,bessi2016social}. This was not the case for Arabic bots. In our data set, bots on average have 81K followers (std = 588K), while humans on average have 7.5K (std = 25.5K). While manually studying accounts, we noticed that suspected bots tend to have a large number of fake followers to amplify their influence and reach. This use of bots (i.e., inflating popularity) has been found to be used by pro-ISIS Twitter accounts \cite{benigni2017online,berger2015isis}. Another special consideration that must be taken into account when analyzing Arabic bots is that some Arabic users use third-party Islamic applications to post Quranic versus automatically on their behalf. This implies that even if some form of automation exists in an account, it doesn't necessarily mean that such an account is a bot.

The result of our regression model shows that Arabic bots can be identified with a high level of accuracy. Our feature analysis showed that bots in our dataset exhibit distinct behaviors and features. Unlike humans, bots tend to not communicate and engage in conversations with other accounts. This characteristic has been found to be true for English bots as well \cite{beskow2018bot}. Significant differences appeared in the distribution of sources used by bots and humans, where we found that bots tend to use third-party applications more often than humans to keep their accounts flowing and active. We also found a significant difference in the distribution of topics discussed by bots and humans. Unlike bots, humans tend to discuss a wider range of topics.

We found linguistic features to be highly discriminatory in detecting Arabic bots. We showed that training the regression model on simple content and linguistic features outperformed Botometer by 20 points in Spearman's rho. This result emphasizes the importance of considering language-specific features in bot detection tasks. Important informative linguistic features include the use of numerics and emojis. We found that bots tend to include in their tweets less emojis and more numbers than humans. Other informative linguistic features include the average length of words and the average number of punctuations marks. Linguistic features especially deceptive language cues have been found to be highly discriminatory for distinguishing English bots as well \cite{addawood2019linguistic}.

The topic of understanding online human behavior has been of a great interest to CSCW/HCI researchers in various contexts such as mental health \cite{de2017gender,ernala2017linguistic}, political polarization \cite{borge2015content,magdy2016isisisnotislam}, and abusive social behaviors \cite{chandrasekharan2017you,soni2018see}. Our findings challenge the assumption often made by such studies that online social media content is always created by humans. We showed that the presence of bots can bias analysis results and disrupt people's online social experience. Platform designers should increase their efforts in combating malicious bots that compromise online democracy. Data scientists should also account for bots in their studies. In particular, Arabic social media studies that are focused on understanding the differences in behaviors and language use between humans and bots can benefit greatly from our bot detection model. For example, a recent study on English Twitter showed how trolls/bots, unlike humans, had been relying on the use of a deceptive/persuasive language in an effort to manipulate the 2016 U.S. elections \citep{addawood2019linguistic}. Having a bot detection tool fitted for Arabic such as the one presented in this paper would make such studies possible in Arabic online social spaces. 

While our results mark an important step toward detecting and understanding Arabic bots, our work has potential limitations. First, despite that our model provides a promising performance on detecting current bots, it needs to be updated regularly with new bot examples in order to capture the continuous and inevitable changes in bot behaviors and characteristics. Second, bots in our study were limited to bots that had a role in spreading religious hatred. It will be worth studying Arabic Twitter bots with a wider range of malicious activities and investigate common features among them. Additionally, it may be useful in future works to investigate a larger set of features (e.g., temporal features and features extracted from followers and friends). It will also be important to investigate the efficacy of combining supervised and unsupervised methods to reduce the high cost of manual labeling without sacrificing much of the accuracy. 

Another important future direction is to investigate the impact of bots on human behavior. In particular, it would be valuable to investigate whether bot-disseminated hateful tweets influence/encourage humans to participate in such discourse either through liking, retweeting, or even authoring new hateful tweets. In a political context, this kind of influence has been shown to exist; Twitter reported that nearly 1.4 million human accounts have made some sort of interaction with content created by bots/trolls during the 2016 U.S. election \cite{TwitterRussian}. If this bot impact on humans can be shown to be effective in the context of hate speech, a more important question would be, can bots be used to decrease online hate speech? In other words, would creating ``good" bots that promote tolerance, acceptance, and diversity values in Arabic social media make an impact on humans? The effect of social norms on prejudice is strongly supported in social psychological literature \cite{paluck2009prejudice,crandall2005conformity}. Studies have also shown that people conform to perceived cultural norm of prejudice and that norms can be influenced \cite{stangor2001changing}. Thus, a more focused question would be, can we leverage bots in online social space to positively influence perceived social norms, which would then make people less prejudiced toward other religious groups? A body of CSCW/HCI research has explored the impact of perceived norms on shaping behavior \cite{chancellor2018norms,pavalanathan2018mind,rho2018fostering}, and thus the potential of bots for positive behavior change is certainly worth investigating in future studies.




\section{CONCLUSION}

In this paper, we have investigated the role of bots in spreading hateful messages on Arabic Twitter. We found that bots were responsible for 11\% of hateful tweets in the hate speech dataset. We further showed that English-trained bot detection models deliver a moderate performance in detecting Arabic bots. Therefore, we developed a more accurate bot detection model trained on various sets of features extracted from 86,346 tweets disseminated by 450 manually-labeled accounts. Finally, we presented a thorough analysis of characteristics and behaviors that distinguish Arabic bots from English Bots and from humans in general. Our results facilitate future Arabic bot detection research in contexts beyond spread of religious hate.

%
\bibliographystyle{ACM-Reference-Format}
\bibliography{sample-base}

%

\end{document}